\documentclass[sigconf]{acmart}
\renewcommand\footnotetextcopyrightpermission[1]{}
\AtBeginDocument{%
  \providecommand\BibTeX{{%
    \normalfont B\kern-0.5em{\scshape i\kern-0.25em b}\kern-0.8em\TeX}}}

\usepackage{algorithm}
\usepackage{algorithmicx}
\usepackage{algpseudocode}
\usepackage{multirow}
\usepackage{xspace}
\usepackage{pifont}

\newcommand{\xhdr}[1]{\vspace{5pt} \noindent {\textbf{#1}}}

\makeatletter
\DeclareRobustCommand\onedot{\futurelet\@let@token\@onedot}
\def\@onedot{\ifx\@let@token.\else.\null\fi\xspace}
\def\eg{\emph{e.g}\onedot} 
\def\ie{\emph{i.e}\onedot}

\theoremstyle{definition}

\newtheorem{defn}{Definition}[section]
\DeclareMathOperator*{\Bernoulli}{Bernoulli}
\DeclareMathOperator*{\poly}{poly}

\DeclareMathOperator*{\var}{var}

\newcommand{\blk}[1]{{\color{black}{#1}}}

\usepackage{xcolor}
\usepackage{subfigure}

\setcopyright{acmcopyright}
\copyrightyear{2021}
\acmYear{2021}
\acmDOI{10.1145/1122445.1122456}

\acmConference[SC '21]{}{November 14--19, 2021}{St. Louis, MO}

\setcopyright{none}
\begin{document}

\title{Overcoming barriers to scalability in variational quantum Monte Carlo}

\author{
{\large \textbf{
Tianchen Zhao$^{1}$, \, Saibal De$^{1}$, \, Brian Chen$^{1}$, \,James Stokes$^{2}$, 
\,Shravan Veerapaneni$^{1,2}$} \vspace{0.05in}\\
   $^{1}${\em Department of Mathematics, University of Michigan, Ann Arbor, MI 48109}\vspace{0.05in} \\
  $^{2}${\em Flatiron Institute, Simons Foundation, New York, NY 10010} \vspace{0.05in}\\
}}

\renewcommand{\shortauthors}{}

\begin{abstract}
The variational quantum Monte Carlo (VQMC) method received significant attention in the recent past because of its ability to overcome the curse of dimensionality inherent in many-body quantum systems. Close parallels exist between VQMC and the emerging hybrid quantum-classical computational paradigm of variational quantum algorithms. VQMC overcomes the curse of dimensionality by performing alternating steps of Monte Carlo sampling from a parametrized quantum state followed by gradient-based optimization.
While VQMC has been applied to solve high-dimensional problems, it is known to be difficult to parallelize, primarily owing to the Markov Chain Monte Carlo (MCMC) sampling step. In this work, we explore the scalability of VQMC when autoregressive models, with exact sampling, are used in place of MCMC. This approach can exploit distributed-memory, shared-memory and/or GPU parallelism in the sampling task without any bottlenecks. In particular, we demonstrate the GPU-scalability of VQMC for solving up to ten-thousand dimensional combinatorial optimization problems.
\end{abstract}


\maketitle

\section{Introduction}
The fact that the state space of a quantum system scales exponentially with the number of its constituents leads to an inevitable curse-of-dimensionality facing the exact simulation of generic quantum many-body systems.

In practice, approximate solutions are sufficient for most purposes and a number of successful variational methods based on the Rayleigh-Ritz principle have been developed, which, given a local Hamiltonian $H$, produce an estimate for the minimal eigenvalue $\lambda_{\rm min}(H)$ and a description of an associated eigenvector. Nevertheless, complexity-theoretic arguments suggest that the curse-of-dimensionality is ultimately unavoidable \cite{aaronson2009quantum} and the investigation of scalable variational algorithms is an active field of research. A particularly promising variational algorithm from the viewpoint of scalability is the variational quantum Monte Carlo (VQMC) \cite{mcmillan-pr65}.

VQMC targets the ground eigenstate by performing alternating steps of Monte Carlo sampling from a high-dimensional quantum state followed by gradient-based optimization. By exploiting neural networks as trial wavefunctions, Carleo and Troyer \cite{carleo2017solving} showed that VQMC can achieve state-of-the-art results for the ground state energies of physically interesting magnetic spin models. Unfortunately, the increased flexibility afforded by neural networks comes at the cost of rendering exact Monte Carlo sampling intractable, which necessitates the use of a Markov Chain Monte Carlo (MCMC) sampling strategy.

However, MCMC sampling limits the scalability of VQMC in two ways: (1) the burn-in process is an inherently sequential task; (2) sampling precise and uncorrelated samples become increasingly difficult for large input dimension. Autoregressive models, in contrast, provide efficient and exact computations for both sampling and density evaluation that are GPU-supported. Recently, autoregressive neural quantum states have been introduced \cite{sharir-prl20}, which has allowed the VQMC to enjoy the advantages that autoregressive models have previously provided in machine learning. Inspired by the ability of autoregressive models to eliminate the reliance of the VQMC on the MCMC, we undertake a parallelization study of autoregressive neural quantum states, thereby improving the time-efficiency and scalability of VQMC.

\section{Background}

In this section, we briefly explain the basics of VQMC, MCMC, and autoregressive models, and state the high-dimensional problems considered.

\subsection{Variational Quantum Monte Carlo}

We consider the problem of determining a minimal eigenpair of a large and sparse random real-symmetric matrix $H$ admitting an efficient description in a sense to be made precise later. 
Moreover, we assume that all off-diagonal entries of $H$ are non-positive so that the ground eigenvector can be chosen to be entry-wise non-negative real vector as a consequence of the Perron-Frobenius theorem. The sparsity assumption  is summarized by the following requirement
\begin{defn}\label{def:sparse}
A real-symmetric matrix $H \in \mathbb{R}^{N\times N}$ is row-$s$ sparse and efficiently row computable if for each row index $x \in [N],$ the list of non-zero entries $\{ (y, H_{xy}) : H_{xy} \neq 0 \}$ is computable in time $O(s)$.
\end{defn}
The specific matrices we will consider are motivated by many-body quantum Hamiltonians.  The size of these matrices is a power of 2, that is, $N = 2^n$, and they have sparsity parameter $s = \poly(n)$ with $n = O(\log N)$. These include as a special case quadratic unconstrained binary optimization (QUBO) problems such as Max-Cut \cite{bravyi2019approximation}.

Given a matrix $H$ satisfying Definition~\ref{def:sparse}, together with differentiable family of trial vectors indexed by $\theta \in \mathbb{R}^d$ described via a function $\psi_\theta : [N] \to \mathbb{R}$ which outputs components of the vector relative to the standard basis $\psi_\theta(x) = \langle e_x , \psi_\theta \rangle$, we define the VQMC learning problem as the following continuous stochastic optimization task,
\begin{equation}\label{e:vmc}
    \min_{\theta \in \mathbb{R}^d} L(\theta) \enspace , \enspace \enspace L(\theta) := \frac{\langle \psi_\theta , H  \psi_\theta \rangle}{\langle \psi_\theta , \psi_\theta \rangle}
    =
    \underset{x \sim \pi_\theta }{\mathbb{E}}\left[\frac{(H  \psi_\theta)(x)}{\psi_\theta(x)}\right] \enspace ,
\end{equation}
where the expectation value is over the  probability distribution
\begin{equation}
    \pi_\theta(x) := \frac{\psi_\theta(x)^2}{ \langle \psi_\theta, \psi_\theta \rangle} \enspace .
\end{equation}
The population objective function \eqref{e:vmc} satisfies the variational inequality $L(\theta) \geq \lambda_{\min}(H)$ and can be concisely expressed as the expectation value of a function  $l_\theta(x)$ (called the local energy for historical reasons),
\begin{equation}
    L(\theta) = \underset{x\sim\pi_\theta}{\mathbb{E}}[l_\theta(x)] \enspace , \enspace \enspace l_\theta(x) := \frac{(H \psi_\theta)(x)}{\psi_\theta(x)} \enspace .
\end{equation}
It follows from Definition~\ref{def:sparse} that each entry of the matrix vector product $H\psi_\theta$ is computable in time $O(s)$ and thus $l_\theta(x)$ is also computable in time $O(s)$ given our sparsity assumption $s = \poly(O(\log N))$. The variance of the stochastic objective under $\pi_\theta$ satisfies the identity.,
\begin{align}
    \var_{x\sim\pi_\theta}\big(l_\theta(x)\big)
    &:= \underset{x\sim\pi_\theta}{\mathbb{E}} \big[ \big(l_\theta(x) - L(\theta)\big)^2 \big] \nonumber \\
    &= \frac{\langle \psi_\theta, H^2 \psi_\theta \rangle}{\langle \psi_\theta, \psi_\theta \rangle} - \left[\frac{\langle \psi_\theta, H \psi_\theta \rangle}{\langle \psi_\theta, \psi_\theta \rangle}\right]^2 \enspace .
\label{eq:stochastic_variance}
\end{align}
Using the Rayleigh-Ritz principle it can be seen that the variance is vanishing if $\psi_\theta$ approaches any eigenvector of $H$.
In practice, the objective function is optimized using stochastic natural gradient descent, also called stochastic reconfiguration (SR) \cite{sorella_aps98}, where the estimators for the gradient and the Fisher information matrix follow from the following population forms,
\begin{align}
    \nabla L(\theta) &= 2\underset{x \sim \pi_\theta}{\mathbb{E}} \big[\big(l_\theta(x) - L(\theta) \big) \nabla_\theta \log |\psi_\theta(x)|\big] \enspace , \nonumber \\
    I(\theta) &= \underset{x \sim \pi_\theta}{\mathbb{E}} \big[\nabla_\theta \log \pi_\theta(x) \otimes \nabla_\theta \log \pi_\theta(x)\big] \enspace .
\end{align}
Typically the normalizing constant $\langle \psi_\theta, \psi_\theta \rangle$ of the probability distribution $\pi_\theta$ is unknown, so the above expectation values are to be approximated using MCMC sampling.

\subsection{Markov Chain Monte Carlo Sampling}

MCMC methods have been developed for sampling from a probability distribution $\pi_\theta$ that is difficult to directly draw \textit{i.i.d.}\ samples from. The canonical Metropolis-Hastings algorithm \cite{hastings1970monte} and its numerous variations, \eg, Gibbs sampling \cite{geman1984stochastic}, Reversible Jump MCMC \cite{green1995reversible} and Hamiltonian Monte Carlo \cite{duane1987hybrid, hoffman2014no}, achieve this by carefully constructing a transition kernel $p(x_{t + 1} | x_t)$ for an ergodic Markov chain whose state distribution limits to the target distribution. Using samples from this Markov chain, we can then compute estimates for the expected values required in VQMC framework
\begin{equation}
    \label{eq:monte-carlo}
    \underset{x \sim \pi_\theta}{\mathbb{E}}[\phi(x)] \approx \overline{\phi}_T = \frac{1}{T} \sum_{t = 1}^T \phi(x_t) \enspace ,
\end{equation}
where $\phi$ represents some deterministic function. Furthermore, these estimates are guaranteed to be asymptotically unbiased by the ergodic theorem.

\begin{figure}[t]
\centering
\includegraphics[width=0.47\textwidth]{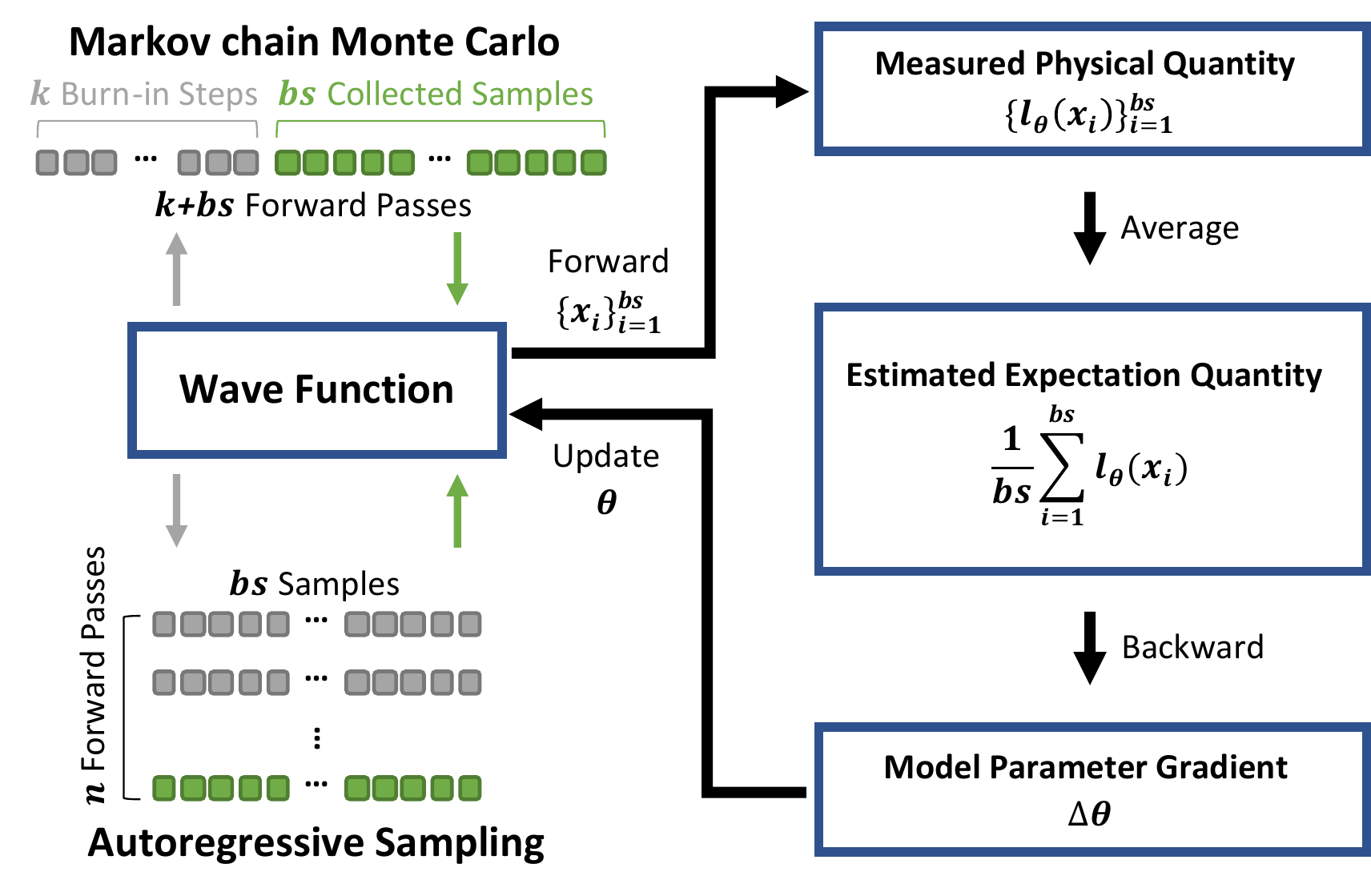}
\caption{Overview of our algorithms, with illustrations of the comparison between Markov chain Monte Carlo sampling (MCMC) and autoregressive sampling (AUTO) on the left, and the VQMC optimization procedure on the right. MCMC sampling involves $k+bs/c$ forward passes, where $k$ is the number of burn-in samples, $c$ is the number of sampling chains ($c = 1$ in the figure) and $\textit{bs}$ is the batch size; AUTO only requires $n$ forward passes to sample \textit{exactly} from the distribution of interest.}
\label{fig:overview}
\end{figure}

\subsection{Autoregressive Models}
Now we discuss the modeling assumptions which enforce normaliztion of the differentiable trial function $\psi_\theta : [N] \to \mathbb{R}$, and thus eliminate the need for MCMC sampling. An elegant method to impose normalization is to make use of an autoregressive assumption, which has recently been generalized to neural network quantum states in \cite{sharir-prl20, hibat2020recurrent}. Since we are targeting a ground eigenvector, which is known to be non-negative, we may assume without loss of generality that $\psi_\theta(x) = \sqrt{\pi_\theta(x)}$, thereby shifting the modeling assumption into the choice of a normalized distribution $\pi_\theta$ satisfying the following condition,
\begin{equation}
    \pi_\theta(x) = \prod_{i=1}^n \pi_i(x_i|x_{i-1},\ldots,x_1) \enspace .
\end{equation}

Many proposals for neural networks satisfying the autoregressive assumption have been put forth. In this work we follow \citet{germain-icml15}, who proposed the masked autoencoder for distribution estimation (MADE) which computes all conditionals in one forward pass using a single network with appropriate masks.
Recall that a single hidden layer autoencoder is described by the following composition of functions,
\begin{align}
    g_1(x) &= \max \{0, W_1x + b_1\} \\
    g_2(x) &= \sigma(W_2g_1(x) + b_2) \enspace ,
\end{align}
and where the rectification and sigmoid functions are applied elementwise. MADE achieves the desired autoregressive assumption by appropriate application of binary masks $M_1$ and $M_2$ to the weight matrices defining the autoencoder, resulting in a MADE layer of the form
\begin{align}
   g_1(x) &= \max\{0, (M_1 \odot W_1) x + b_1 \} \nonumber \\
   g_2(x) &= \sigma\left((M_2 \odot W_2) g_1(x) + b_2\right) \enspace ,
\end{align}
where $\odot$ denotes elementwise multiplication.

In Figure~\ref{fig:overview}, we compare the sampling procedures between MCMC and AUTO (as described in Algorithm~\ref{algo:autoregressive_sampling}).
MCMC involves $k+bs/c$ forward passes, where $c$ is the number of sampling chains and $\textit{bs}$ is the batch size. Although the number of forward passes can be reduced by increasing the number of chains, the number of burn-in iterations $k$ required for convergence is undetermined and cannot be parallelized. On the other hand, AUTO only requires $n$ forward passes to sample exactly from the distribution of interest.

\begin{algorithm}[t!]
\caption{Autoregressive Sampling~\cite{germain-icml15} (Batch Size 1)}
\begin{algorithmic}
    \State \textbf{Input}: Randomly initialized state $x^0$ of size $n$
    \State \textbf{Output}: Sampled state $x^*$ of size $n$
    \For{$i$th out of $n$ iterations}
        \State Compute $p(x_i|x^{i-1}_{1:i-1})$ with a forward pass
        \State Sample $y_i\in\{\pm 1\}$ with $p(\cdot|x^{i-1}_{1:i-1})$
        \State Get $x^i$ by updating $x^{i-1}_i$ with $y_i$
    \EndFor
    \State Set $x^*=x^n$
\end{algorithmic}
\label{algo:autoregressive_sampling}
\end{algorithm}

\subsection{Quantum Hamiltonians and QUBO Problems}
In this paper, we consider a family of
matrices motivated by quantum physics,
which are parametrized by $O(\poly(n))$ real parameters $\alpha_i,\beta_i,\beta_{ij} \in \mathbb{R}$ as follows,
\begin{equation}
    H = -\sum_{1 \leq i \leq n} \big(\alpha_i X_i + \beta_i Z_i\big) - \sum_{1 \leq i < j \leq n} \beta_{ij} Z_i Z_j \enspace ,
\label{eq:hamiltonian}
\end{equation}
where $X_i := I^{\otimes (i - 1)} \otimes X \otimes I^{\otimes(n-i)}$ and $Z_i := I^{\otimes (i - 1)} \otimes Z \otimes I^{\otimes(n-i)}$ are $2^n\times 2^n$ matrices defined in terms of the following elementary $2\times 2$ matrices,
\begin{equation}
    I =
    \begin{bmatrix}
    1 & 0 \\
    0 & 1
    \end{bmatrix}
    \enspace ,
    \quad \quad \quad
    Z =
    \begin{bmatrix}
    1 & 0 \\
    0 & -1
    \end{bmatrix}
    \enspace ,
    \quad \quad \quad
    X =
    \begin{bmatrix}
    0 & 1 \\
    1 & 0
    \end{bmatrix}
    \enspace .
\end{equation}
It is easily verified that $H$ meets the conditions of definition \ref{def:sparse} with sparsity parameter $s=n$.
In terms of the binary representation of the row index $x = 2^{n-1}x_1 \cdots 2^0x_n$ and the column index $y = 2^{n-1} y_1 \cdots 2^0 y_n$, the matrix entries of $H$ are given by
\begin{align}\label{e:tim}
    H_{xy} =& -\sum_{1 \leq i \leq n} \big(\alpha_i \delta_{x_1 y_1} \cdots \delta_{\neg x_iy_i}\cdots \delta_{x_ny_n} + \beta_i (1-2x_i)\big) \nonumber \\
    & -\delta_{xy}\sum_{1 \leq i \leq j \leq n} \beta_{ij} (1-2 x_i)(1-2x_j)
\end{align}
and $\neg x_i$ denotes logical negation of $x_i \in \{0,1\}$.
For simplicity we imposed $\alpha_i \geq 0$ to ensure that the ground eigenvector can be chosen to be a non-negative vector as a consequence of the Perron-Frobenius theorem.

In the special case where $\alpha_i = \beta_i = 0 $ and  $\beta_{ij} = \frac{1}{4}L_{ij}$ where $L$ is the adjacency matrix of an undirected graph $G = (V, E)$ of size $|V|=n$, the ground state problem coincides with the Max-Cut problem, and thus VQMC can be employed as a heuristic for approximate combinatorial optimization \cite{gomes2019classical, zhao2020natural}, which is equivalent to natural evolution strategies \cite{zhao2020natural}.

\section{Related Work}
The idea of utilizing neural network quantum states to overcome the curse of dimensionality in high-dimensional VQMC simulations was first introduced by Carleo and Troyer \cite{carleo2017solving}, who concentrated on restricted Boltzmann machines (RBMs) applied to two-dimensional quantum spin models. Sharir \emph{et al.} \cite{sharir-prl20,sharir-git20} introduced neural network quantum states based on the autoregressive assumption inspired by PixelCNN \cite{oord2016pixelcnn} and demonstrated significant improvement in performance compared to RBMs. The autoregressive assumption was subsequently explored in VQMC using recurrent neural wavefunctions \cite{hibat2020recurrent}. Autoregressive models have also been used to solve statistical mechanics models in \cite{wu2019solving} Since our focus is on the scalability of VQMC, particularly in situations where MCMC is expected to struggle, unlike \cite{carleo2017solving, sharir-prl20, hibat2020recurrent} we consider non-geometrically local Hamiltonians without an underlying lattice structure. This also contrasts with the work of \cite{misawa-cpc19}, who considered parallelization of VQMC using MCMC sampling but assuming geometric locality.
 It was recently shown \cite{gomes2019classical, zhao2020natural} that techniques from quantum VQMC literature \citep{carleo2017solving} can be adapted for approximately solving combinatorial optimization problems.

\citet{larochelle2011neural} proposed neural autoregressive distribution estimator (NADE) as feed-forward architectures. MADE~\cite{germain-icml15} improves the efficiency of models with minor additional cost for simple masking operations. For probabilistic generative models, unnormalized models such as RBM rely on approximate sampling procedures like MCMC, whose convergence time remains undetermined, which often results in the generation of highly correlated samples and deterioration in performance. Such sampling approximations can be avoided by using autoregressive models~\cite{bengio2000modeling} that estimate the joint distribution by decomposing it into a product of conditionals by the probability chain rule, making both the density estimation and generation process tractable.
\citet{kingma-neurips16} used autoregressive models as a form of normalizing flow~\cite{kobyzev-tpami20}.

\section{Algorithm Parallelization}

Unlike standard Monte Carlo methods, MCMC cannot be parallelized easily. The fundamental limitation is easily seen: to generate a sample $x_{t + 1}$ from a Markov chain, we need to sample the transition kernel $p(\cdot | x_t)$, which requires knowledge of the immediate past state $x_t$. This sequential nature of the sampling immediately precludes any direct attempt at parallelizing the sampling process.

We could attempt to initialize multiple independent sampling chains; indeed, this is one of the standard approaches often implemented in Bayesian inference frameworks. But when sampling a high-dimensional distribution using random walk Metropolis-Hastings, it typically takes a very long time for the random walk to explore the parameter space. This significantly slows down the convergence of the estimates \eqref{eq:monte-carlo} to the true expectation value; furthermore, it is very difficult to determine \textit{a priori} how many samples will be required for this convergence within a specified tolerance. In practice, MCMC first discards a pre-determined number of samples in each of the independent chains to avoid the transient Markov transitions (a.k.a.\ burn-in) and down-samples the remainder by selecting samples at regular intervals to reduce correlations (a.k.a.\ thinning). Any expectations are then computed based on this smaller set of selected samples. Improper choice of these parameters can severely degrade the quality of the generated estimates. Furthermore, they also reduce the parallel efficiency; suppose $k$ samples are discarded as burn-in and every $j$-th samples are selected during thinning. Then constructing $n$ samples on each of $L$ independent computing units will lead to a parallel efficiency of
\begin{equation}
    \frac{k + (n L - 1) j + 1}{k + (n - 1) j + 1} = 1 + \frac{n j}{k + (n - 1) j + 1} (L - 1) = a + b L
\end{equation}
for some $a$ and $b$ depending on $k$, $j$ and $n$. Note that this calculation is solely focused on the sampling task, and therefore does not take into account any communications that might be necessary between the computing units for obtaining the final result. Even then, as the number of burn-in samples $k$ is increased, the slope $b$ decays from 1 towards 0 ($b = 1$ is indicative of optimal scaling).

On the other hand, an autoregressive model (AUTO) can generate exact samples from the target distribution. Although the implementation of AUTO has a sequential nature that scales linearly with the input dimension, it can generate independent samples from the target distribution by transforming \textit{i.i.d.}\ samples from a simple distribution (\eg Gaussian). This step is easily parallelized: as long as we have identical copies of the autoregressive model in a number of computing units (\eg GPUs), we can construct independent samples in parallel. Communication between the computing units is necessary only when we need to update the parameters of the neural network, \eg during a stochastic gradient descent update.

Our model consists of fully connected weight matrices; therefore as we scale up the problem size, the bottleneck for our algorithm is the memory usage.
For example, assuming a GPU can only store models with up to 10M parameters, we can set the size of the hidden layer to 500 at maximum when solving a problem with 10K input dimensions. This limitation can be addressed along with two complementary but independent avenues:
\begin{enumerate}
    \item \textbf{Model Parallelization:} Distribute the model parameters across computing units, so that each unit needs to store and update a small part of the model.

    \item \textbf{Sampling Parallelization:} Use identical copies of the model across the computing units to generate only a few samples per unit, and combine the independent samples from all these units to construct an accurate expectation estimate.
\end{enumerate}
The communication pattern between the computing units in model parallelization is intimately linked with the choice of the autoregressive neural network while the sampling parallelization is model agnostic.

In this work, we restrict our attention to only parallelizing the sampling step. Consider a quantum Hamiltonian of size $N = 2^n$ and an autoregressive model with two hidden layers of size $h$. Given a total number of $L$ computing units/GPUs and a mini-batch size of $mbs$ samples to be drawn on each GPU, we end up with an effective batch size of $bs = L \times mbs$. Locally, each process first generates $mbs$ samples, then computes the physical measurements with the samples, and finally uses backpropagation to get the gradient of the model parameters. These local gradient vectors have length $d = 2 h n + h + n$, which are averaged over the GPUs using a parallel reduction. Each GPU then updates its own model parameters locally.

The computation complexity can be estimated as follows: during the local sampling process on each GPU, the algorithm involves $n$ forward passes for sampling, and a fixed number of forward passes for physical quantity measurements. The dominant cost of each forward pass is multiplication by $h \times n$ and $n \times h$ matrices, both $O(h n)$; this leads to a total computational cost of $O(h n^2 \times mbs)$ flops per GPU. Computing the average gradient over GPUs using parallel reduction costs further $O(h n)$ flops, and involves communication of $O(h n)$ floating point numbers. Clearly, the parallel efficiency is given by
\begin{equation}
    \frac{O(h n^2 \times bs)}{O(h n^2 \times mbs) + O(h n)} = \frac{O(h n^2 \times L \times mbs)}{O(h n^2 \times mbs) + O(h n)}
\end{equation}
Since the constants in the $O(h n^2 \times L \times mbs)$ and the $O(h n^2 \times mbs)$ are the same, this ratio is approximately $L$ when $n$ or $mbs$ are large.


\section{Experimental Results}

\begin{table*}[t]
\caption{Training time (measured in seconds) comparison on TIM for 300 training iterations with one GPU. Our MCMC settings are introduced in Section~\ref{sec:experimental_setup}. The running time of MADE\&AUTO scales roughly linearly with respect to the number of dimensions, due to the sequential nature of its sampling procedure, but significantly outperforms RBM\&MCMC in practice.}
\label{tbl:benchmark_running_time}
\begin{tabular}{lccccccc}
\toprule
\multirow{2}{*}{Model} & \multirow{2}{*}{Optimizer} & \multirow{2}{*}{Sampler} & \multicolumn{5}{c}{\# of Dimensions} \\
\cmidrule(lr){4-8}
& & & 20 & 50 & 100 & 200 & 500 \\
\midrule
RBM & ADAM & MCMC & 135.64 & 154.25 & 189.91 & 249.40 & 456.68 \\
MADE & ADAM & AUTO & 2.85 & 5.74 & 10.63 & 20.45 & 49.62 \\
\toprule
\end{tabular}
\end{table*}

\begin{figure*}[t]
\centering
\includegraphics[width=0.98\textwidth]{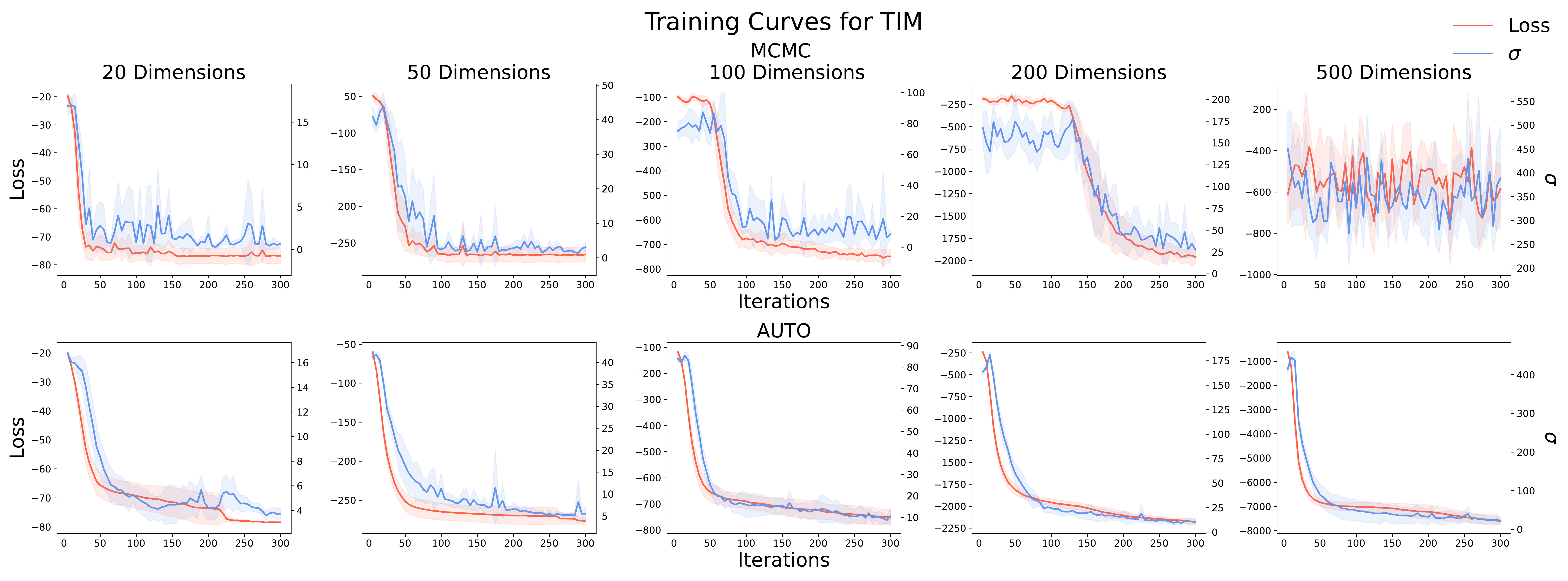}
\caption{Training curves for TIM, where the red curves refer to the training loss/energy, and the blue curves refer to the standard deviation of the stochastic objective, which should be zero when the wave function converges to the exact ground-state. By fixing the learning rate and the total number of training iterations, it becomes more difficult for RBM\&MCMC to converge as the problem size grows, due to the inaccurate estimation of the population energy by the low-quality MCMC samples. The training of MADE\&AUTO is stable across all problems.}
\label{fig:curves}
\end{figure*}

This section contains an extensive evaluation of our
approach. We first compare AUTO sampling and MCMC sampling in Section~\ref{sec:MCMC_vsv_AUTO}, where the advantage of AUTO in terms of computational efficiency becomes clear for problems of higher dimensions. The convergence performance is shown in Section~\ref{sec:convergence}. Our algorithm is competitive against the state-of-the-art SDP solvers for small/medium scale Max-Cut problems. In Section~\ref{sec:scalability}, we demonstrate the scalability of our technology by solving large-scale problems up to 10K dimensions. We achieved near-optimal weak scaling, and the convergence of our model improves as we increase the effective training batch size.

\subsection{Experimental Setup}
\label{sec:experimental_setup}

In this paper, we evaluate VQMC using two non-geometrically local Hamiltonians: the Max-Cut and the transverse field Ising model (TIM) model.
In the case of Max-Cut, the adjacency matrix was chosen by forming the $n \times n$ matrix $(B+B^T)/2$ with $B_{ij} \sim \Bernoulli(0.5)$ sampled once and fixed,  followed by rounding and setting diagonal entries to zero. The second example is a disordered quantum system referred to as transverse field Ising model model, whose Hamiltonian is of the form \eqref{e:tim} with $\beta_i, \beta_{ij} \sim U(-1,1)$ and $\alpha_i \sim U(0, 1)$ sampled once and fixed.

For Max-Cut, we compare our approach against VQMC with MCMC sampling~\cite{gomes2019classical,zhao2020natural}, as well as the semidefinite programming (SDP) relaxation approximation algorithms including
Goemans-Williamson Algorithm~\cite{goemans1995improved} and the Burer-Monteiro reformulation with the Riemannian Trust-Region method~\citep{absil-fcm07}. As an additional baseline, each model is also trained using the SR method. We benchmark the running time and converged energy of our model on TIM in our scalability experiments.

\xhdr{Model architecture}

Network architecture is chosen to be MADE and is compared against RBM, proposed by \citet{carleo2017solving}, taking the one-dimensional state as input and outputs the logarithmic probability amplitude.

The structure of MADE is as follows
\begin{align}
    \textit{Input} &\xrightarrow{[\textit{bs},n]} \texttt{MaskedFC1} \xrightarrow{[\textit{bs},h]} \texttt{ReLU} \nonumber \\ &\xrightarrow{[\textit{bs},h]} \texttt{MaskedFC2} \xrightarrow{[\textit{bs},n]} \texttt{Sigmoid} \xrightarrow{[\textit{bs},n]} \textit{Output}, \nonumber
\end{align}
and the structure of RBM is
\begin{align}
    \textit{Input} &\xrightarrow{[\textit{bs},n]} \texttt{FC}_{n,h} \xrightarrow{[\textit{bs},h]} \texttt{Lncoshsum} \xrightarrow{[\textit{bs}]} \textit{Output}1 \nonumber\\
    & \xrightarrow{[\textit{bs},n]} \texttt{FC}_{n,1} \xrightarrow{[\textit{bs}]} \texttt{Add}\;\textit{Output}1 \xrightarrow{[\textit{bs}]} \textit{Output}. \nonumber
\end{align}
Here $\textit{bs}$ is the batch size and $n$ is the number of dimensions. $\texttt{FC}_{a,b}$ is a fully connected layer with input size $a$ and output size $b$; and $\texttt{MaskedFC}$ is the masked version of $\texttt{FC}$, to remove the connections in the computational path of MADE. \texttt{Lncoshsum} refers to a series of linear and non-linear operations involving: 1) taking natural logarithm for each entry of the input tensor; 2) taking hyperbolic cosine for each entry of the input tensor; 3) summation over the last dimension of the input tensor. The size of the tensor being passed to the next operator is indicated above the right arrows.

For large-scale problems with high dimensional input size $n$, we need to choose a proper latent size $h$ to balance between the memory usage and the capacity of the model. In our experiments, we set $h=5(\log n)^2$ as the hidden layer size for MADE and $h=n$ as the number of hidden units for RBM.

\begin{table*}[t]

\caption{Optimized objective (maximize cut number for Max-Cut, minimize ground state energy for TIM) values for different problem sizes and different optimizers, averaged over 5 runs with different random seeds. The first three rows in the Max-Cut section consist of results from running classical algorithms and serve as benchmarks. For the rest of the rows in the table, the batch size is fixed to be 1024. We note that MADE\&AUTO achieves satisfactory performance in the sense that it's directly comparable with the SDP solvers on Max-Cut. On the other hand, RBM\&MCMC takes longer to converge as the problem size grows, whereas the convergence of MADE\&AUTO remains stable.}
\label{tbl:benchmark_convergence}
\begin{tabular}{ccccccccc}
\toprule
\multirow{2}{*}{Problem} & \multirow{2}{*}{Model} & \multirow{2}{*}{Sampler} & \multirow{2}{*}{Optimizer} &\multicolumn{5}{c}{\# of Dimensions} \\
\cmidrule(lr){5-9}
& & & & 20 & 50 & 100 & 200 & 500 \\
\midrule
\multirow{9}{*}{Max-Cut} & \multicolumn{3}{l}{Classical: Random} & 27.2 $\pm$ 2.2 & 150.4 $\pm$ 5.8 & 610.4 $\pm$ 11.6 & 2495.8 $\pm$ 42.8 & 15696.0 $\pm$ 16.8 \\
& \multicolumn{3}{l}{Classical: Goemans-Williamson} & 41.4 $\pm$ 2.0 & 194.2 $\pm$ 2.3 & 741.0 $\pm$ 11.1 & 2881.6 $\pm$ 14.4 & 17242.4 $\pm$ 37.3 \\
& \multicolumn{3}{l}{Classical: Burer–Monteiro} & \textbf{43.0} $\pm$ 0.0 & \textbf{200.0} $\pm$ 0.0 & 754.0 $\pm$ 3.0 & \textbf{2928.0} $\pm$ 3.7 & \textbf{17416.0} $\pm$ 23.13 \\
\cmidrule(lr){2-9}
& \multirow{3}{*}{RBM} & \multirow{3}{*}{MCMC} & SGD & 41.4 $\pm$ 1.5 & 192.0 $\pm$ 3.3 & 733.8 $\pm$ 13.0 & 2825.6 $\pm$ 5.5 & 15945.6 $\pm$ 44.2 \\
& & & ADAM & 40.6 $\pm$ 1.6 & 190.2$ \pm$ 2.7 & 719.8 $\pm$ 6.6 & 2777.6 $\pm$ 14.2 & 16576.0 $\pm$ 30.9 \\
& & & SGD+SR & \textbf{43.0} $\pm$ 0.0 & 198.8 $\pm$ 1.5 & 758.0 $\pm$ 1.1 & 2898.0 $\pm$ 22.0 & 15956.8 $\pm$ 29.9 \\
\cmidrule(lr){2-9}
& \multirow{3}{*}{MADE} & \multirow{3}{*}{AUTO} & SGD & 42.6 $\pm$ 0.4 & 192.0 $\pm$ 2.4 & 742.2 $\pm$ 5.9 & 2846.0 $\pm$ 4.8 & 16880.0 $\pm$ 73.6 \\
& & & ADAM & 42.4 $\pm$ 0.8 & 193.8 $\pm$ 3.1 & 733.8 $\pm$ 9.1 & 2847.8 $\pm$ 12.1 & 17006.6 $\pm$ 23.0 \\
& & & SGD+SR & \textbf{43.0} $\pm$ 0.0 & \textbf{200.0} $\pm$ 1.5 & \textbf{758.4} $\pm$ 6.5 & 2909.2 $\pm$ 3.1 & 17176.6 $\pm$ 30.5 \\
\midrule
\multirow{6}{*}{TIM} & \multirow{3}{*}{MCMC} & \multirow{3}{*}{RBM} & SGD & -80.22 $\pm$ 2.79 & -270.65 $\pm$ 9.64 & -762.11 $\pm$ 28.58 & -1981.17 $\pm$ 72.19 & -976.25 $\pm$ 119.43 \\
& & & ADAM & -80.38 $\pm$ 2.42 & -265.47 $\pm$ 8.21 & -756.33 $\pm$ 16.73 & -2216.45 $\pm$ 31.95 & -924.53 $\pm$ 121.10 \\
& & & SGD+SR & -80.70 $\pm$ 2.10 & \textbf{-282.02} $\pm$ 8.37 & -764.74 $\pm$ 14.67 & -2234.23 $\pm$ 36.72 & -1046.40 $\pm$ 334.50 \\
\cmidrule(lr){2-9}
& \multirow{3}{*}{MADE} & \multirow{3}{*}{AUTO} & SGD & -80.30 $\pm$ 0.01 & -281.18 $\pm$ 5.51 & -767.88 $\pm$ 13.45 & -1872.16 $\pm$ 41.89 & -6773.97 $\pm$ 233.19 \\
& & & ADAM & -80.48 $\pm$ 0.18 & -277.11 $\pm$ 4.48 & -771.11 $\pm$ 17.06 & -2181.31 $\pm$ 33.39 & -7597.37 $\pm$ 171.25 \\
& & & SGD+SR & \textbf{-81.25} $\pm$ 0.07 & -277.23 $\pm$ 9.96 & \textbf{-812.33} $\pm$ 12.55 & \textbf{-2252.12} $\pm$ 84.00 & \textbf{-8673.27} $\pm$ 304.45 \\
\toprule
\end{tabular}
\end{table*}

\xhdr{Training}

All models are trained for 300 iterations.
In our single-GPU experiments, at each iteration, the model is updated with a batch of 1024 training samples. For evaluation, we draw a batch of 1024 testing samples from trained model, and report their mean energy.
Two base optimizers are considered: stochastic gradient descent (SGD) with learning rate 0.1 or ADAM with learning rate 0.01, where the latter is our default optimizer. In addition, we provide additional results on models trained using the SR~\cite{sorella_aps98} method for performance comparison. The SR optimization was performed using a regularization parameter $\lambda=0.001$ and a learning rate $0.1$. No learning rate scheduler is applied. For scalability experiments, each GPU is distributed with a constant mini-batch size $mbs$, and the effective batch size is $mbs \times L$, where $L$ is the total number of GPUs available.

Our MCMC sampler is the random walk Metropolis–Hastings algorithm, running with two chains. We expect that it takes more effort for MCMC to converge for large-scale problems. Therefore, for each chain, we set heuristically the burn-in iterations $k$ to scale linearly with respect to the input dimension $n$, \ie, $k=3n+100$.

Throughout the experiments, the timing benchmarks are performed on NVIDIA Tesla V100 GPUs, with 32GB of memory for each.

\subsection{MCMC vs.\ AUTO: Runtime}
\label{sec:MCMC_vsv_AUTO}

Despite the sequential nature of both MCMC and AUTO sampling, in practice, AUTO sampling can be operated with GPU in a straightforward fashion and exhibit superior running time efficiency. Our results on the running time comparison is shown in Table~\ref{tbl:benchmark_running_time}.

The running time of RBM\&MCMC scales with the total number of iterations in each chain, which includes a fixed number of burn-in iterations that cannot be parallelized. In our setting, we set the number of chains to be 2, and burn-in iterations $k$ that grows linearly with respect to the input dimension $n$. In principle, the running time of MCMC can be reduced further by increasing the number of chains or choosing a smaller $k$. However, a more severe problem of MCMC lies in the fact that the distribution of the samples generated by MCMC only converges to the distribution of interest asymptotically. As the input dimension increases, it becomes more difficult for the random walk Metropolis–Hastings algorithm to converge, which can potentially affect the quality of generated samples if $k$ is not properly chosen. The running time of MADE\&AUTO is dominated by the sampling time that scales linearly with respect to the input dimension $n$, which significantly outperforms its RBM\&MCMC counterpart. More importantly, for AUTO, we know exactly the computational complexity needed to get correct samples from the distribution of interest, as opposed to MCMC that requires undetermined number of iterations to converge.

The corresponding training curves are shown in Figure~\ref{fig:curves}, where the red curves refer to the training loss/energy, and the blue curves refer to the standard deviation of the stochastic objective, which approaches zero as the wave function converges to the exact ground-state, as discussed in Eq.~\ref{eq:stochastic_variance}. RBM\&MCMC converges reasonably well on small-scale problems, but has more difficulty to converge as the problem scales up. On the other hand, our model converges rapidly and stably to low energy across problems of different scales. This observation motivates us to attempt to solve problems of even higher dimensions.

\begin{figure*}[t]
\centering
\includegraphics[width=0.9\textwidth]{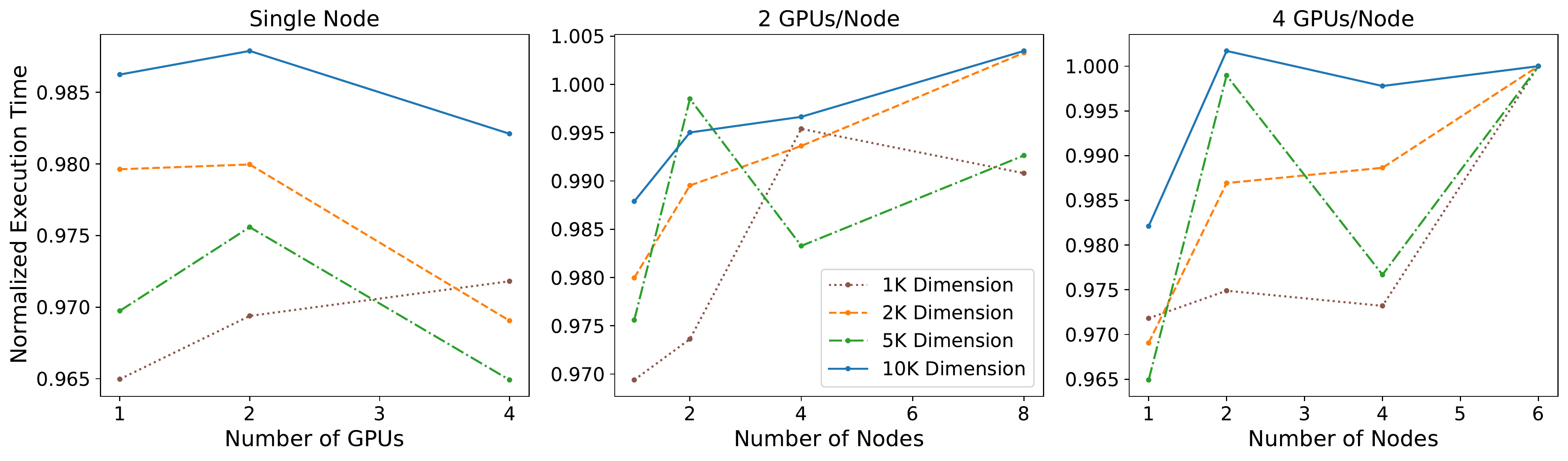}
\caption{Sampling times for the TIM problem in 1K, 2K, 5K and 10K dimensions with mini-batch sizes \textit{mbs} = 512, 128, 16 and 4 samples per GPU, respectively. The minibatch sizes were chosen to saturate GPU memory per problem dimension. All times are normalized by the execution time of the largest GPU configuration ($6 \times 4$) for each dimension. Note that the normalized executions times are all close to $1$, indicative of near-optimal weak scaling.}
\label{fig:weak_scalability}
\end{figure*}

\subsection{MCMC vs.\ AUTO: Convergence Study}
\label{sec:convergence}

The convergence result of our model on the Max-Cut problems is shown in Table~\ref{tbl:benchmark_convergence}, where we compare MADE\&AUTO against the state-of-the-art SDP relaxation approximation algorithms developed in the past decades, as well as VQMC with RBM\&MCMC.

Random Cut algorithm is a simple randomized 0.5-approximation algorithm that randomly assigns each node to a partition. \citet{goemans1995improved} improved the performance ratio from 0.5 to at least 0.87856, by making use of the semidefinite programming (SDP) relaxation of the original integer quadratic program. \citet{burer-mp01} reformulated the SDP for Max-Cut into a non-convex problem, with the benefit of having a lower dimension and no conic constraint. The implementation of Goemans-Williamson Algorithm used the \texttt{CVXPY}~\citep{cvxpy,cvxpy_rewriting} package and the Burer-Monteiro reformulation with the Riemannian Trust-Region method~\citep{absil-fcm07} used \texttt{Manopt} toolbox~\citep{boumal-jml14}, which essentially implements the optimization algorithm proposed by \citep{journee-siam10}.

For evaluation, we constructed a problem instance for each Hamiltonian size $n \in \{20, 50, 100, 200, 500\}$ by randomly generating parameters defined in Eq.~\ref{eq:hamiltonian}. For each problem instance, each algorithm was executed 5 times using 5 random seeds. In Table~\ref{tbl:benchmark_convergence}, we report the averaged result over problem instances of different sizes.

In general, MADE\&AUTO slightly outperforms RBM\&MCMC on small-scale problems, and the latter fails to converge for problems of input dimension 500, due to our constraint on the number of training iterations.

The natural gradient descent \citep{amari1998natural,sorella_aps98} proved essential for converging to a good local optimum. We apply the SR to both VQMC methods and observe similar improvements: optimizers equipped with SR are consistently improved over all architectures. On the other hand, the performance of our algorithm with SR is competitive against the state-of-the-art SDP solvers on Max-Cut problems.

\begin{figure}[t]
\centering
\includegraphics[width=0.47\textwidth]{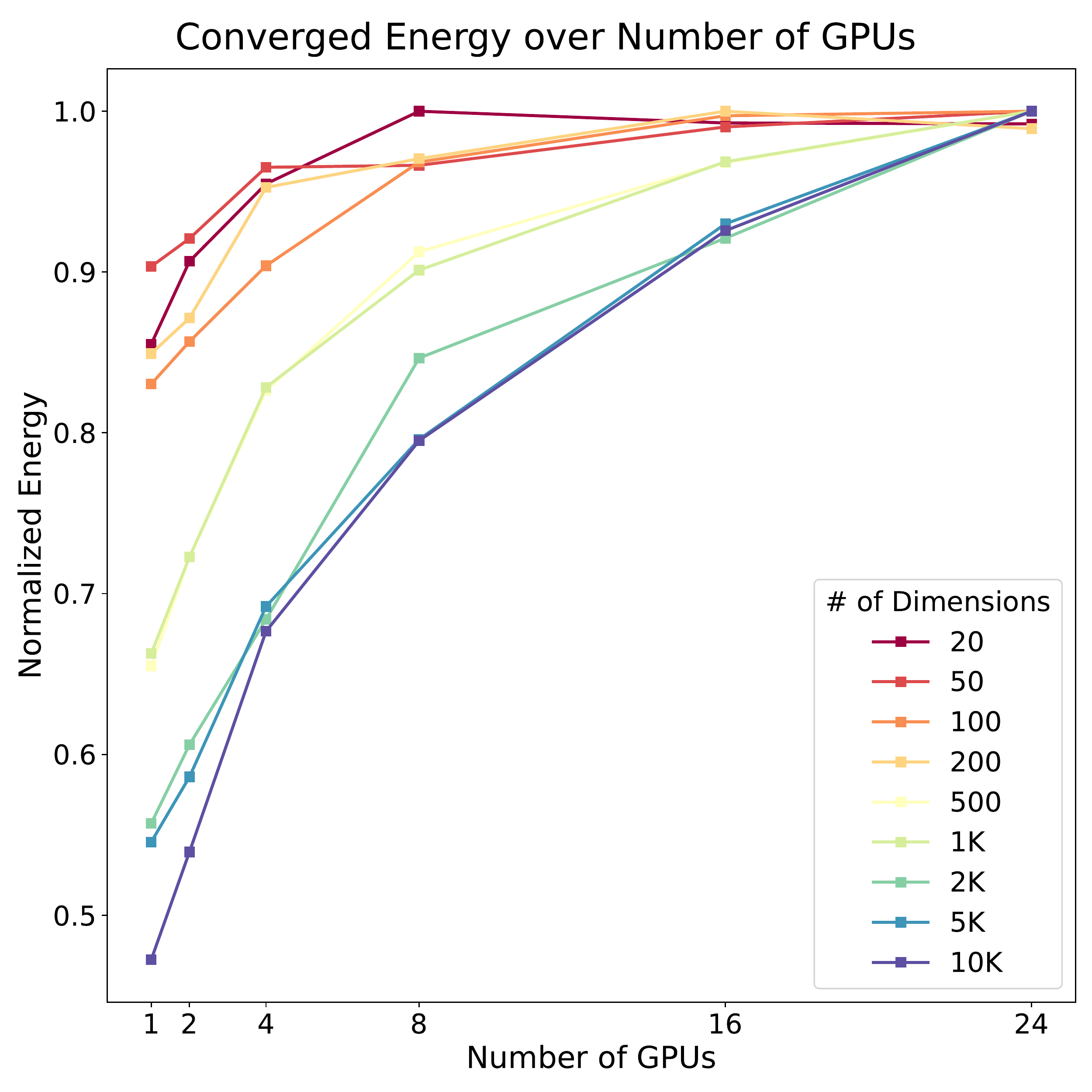}
\caption{Normalized converged energy for TIM problems of different sizes. Each GPU is distributed a batch size of 4; the total effective batch size equals 4 times the total number of GPUs used. The energy is normalized for each problem size (the values from each curve are divided by the one with the largest magnitude among them). The converged energy improves as the total number of GPUs (effective batch size) increases. The improvement saturates for smaller problems, which also implies that a larger batch size is required for larger problems.}
\label{fig:scalability_performance}
\end{figure}

\begin{table*}[t]
\small
\caption{Ablation study on the latent size. We train the models with ADAM on Max-Cut problems. $n$ is the graph size. Optimal performance is obtained under a proper choice of latent size $h$; MADE falls off if we push GPU to its computational limits.}
\label{tbl:ablation_latent_dim}
\begin{tabular}{ccccccccc|ccccc}
\toprule
\multirow{3}{*}{Model} & \multirow{3}{*}{$n$} & \multicolumn{12}{c}{Latent size $h$} \\
& & \multicolumn{7}{c}{Cut table} & \multicolumn{5}{c}{Time table} \\
\cmidrule(lr){3-14}
& & $(\log n)^2$ & $3(\log n)^2$ & \blk{$5(\log n)^2$} & \blk{$n$} & $n$ & $5n$ & $n^2$ & $(\log n)^2$ & $3(\log n)^2$ & $n$ & $5n$ & $n^2$ \\
\midrule
\multirow{4}{*}{MADE}     &  50 & 191. & 192.8 & \blk{193.8} & \blk{-} & 195. & 194.6 & 195.
                               & 7.22 & 7.19 & 7.24 & 7.42 & 7.41 \\
                         & 100 & 735.8 & 737.2 & \blk{733.8} & \blk{-} & 734.2 & 731.2 & 726.2
                               & 13.43 & 13.49 & 13.48 & 13.90 & 13.96 \\
                         & 200 & 2832.8 & 2846.4 & \blk{2847.8} & \blk{-} & 2848.6 & 2821.4 & 2779.
                               & 26.49 & 25.78 & 26.07 & 26.85 & 57.19 \\
                         & 500 & 16905.4 & 17039.6 & \blk{17006.6} & \blk{-} & 16973.8 & 16872.8 & 16311.4
                               & 64.81 & 66.48 & 67.79 & 105.97 & 1426.92 \\
\midrule
\multirow{4}{*}{RBM}    &  50 & 193. & 194.8 & \blk{-} & \blk{190.2} & 192. & 192.2 & 191.4
                               & 151.07 & 151.49 & 150.72 & 150.71 & 152.68 \\
                         & 100 & 721. & 734.2 & \blk{-} & \blk{719.8} & 730.2 & 711. & 705.2
                               & 181.11 & 180.30 & 180.47 & 182.15 & 183.62 \\
                         & 200 & 2786.2 & 2810.8 & \blk{-} & \blk{2777.6} & 2779.6 & 2765.6 & 2747.4
                               & 242.95 & 241.05 & 243.24 & 243.91 & 246.05 \\
                         & 500 & 16568.8 & 16530. & \blk{-} & \blk{16576.0} & 16652.6 & 16577.2 & 16543.
                               & 427.23 & 429.07 & 432.39 & 428.17 & 510.02 \\
\toprule
\end{tabular}
\end{table*}

\begin{table*}[t]
\small
\caption{Ablation study on the MCMC sampling scheme. We train the RBM with ADAM on Max-Cut problems. $n$ is the graph size; $\{n,10n\}$ and $\{\text{\texttimes}2, \text{\texttimes}5, \text{\texttimes}10\}$ are from Scheme 1 and Scheme 2, respectively.}
\label{tbl:ablation_MCMC_sampling}
\begin{tabular}{cccccccc|ccccc}
\toprule
\multirow{3}{*}{Model} & \multirow{3}{*}{$n$} & \multicolumn{11}{c}{Sampling scheme} \\
& & \multicolumn{6}{c}{Cut table} & \multicolumn{5}{c}{Time table} \\
\cmidrule(lr){3-13}
& & $n$ & $3n$+$100$ & $10n$ & \texttimes$2$ & \texttimes$5$ & \texttimes$10$ & $n$ & $10n$ & \texttimes$2$ & \texttimes$5$ & \texttimes$10$ \\
\midrule
\multirow{4}{*}{MCMC}     &  50 & 190.8 & \blk{190.2} & 193.8 & 191.6 & 192.6 & 192.8
                               & 110.44 & 197.02 & 199.64 & 500.02 & 1004.96 \\
                         & 100 & 700.2 & \blk{719.8} & 733. & 706.8 & 720. & 729.8
                               & 124.01 & 296.83 & 201.52 & 507.65 & 1011.51 \\
                         & 200 & 2674.8 & \blk{2777.6} & 2795.4 & 2670.4 & 2720.6 & 2736.8
                               & 143.76 & 492.31 & 206.91 & 514.80 & 1023.43 \\
                         & 500 & 16205. & \blk{16576.0} & 16626.6 & 16022.2 & 16066.6 & 16156.6
                               & 212.86 & 1103.18 & 207.43 & 508.43 & 1021.21 \\
\toprule
\end{tabular}
\end{table*}

\subsection{AUTO: Multi-GPU Scalability}
\label{sec:scalability}

By distributing the sampling task across multiple GPUs, our method can extend to large-scale problems (with input dimensions up to 10K) by reducing the mini-batch size $\textit{mbs}$ distributed to each GPU. The effective batch size depends on both $\textit{mbs}$ and the number of GPUs $L$ available for training.

In Figure~\ref{fig:weak_scalability}, we plot the normalized execution times for the 1K, 5K and 10K dimensional TIM problems as we vary the number of GPUs and the GPU distribution across nodes. We choose the minibatch sizes assigned to each GPU depending on the dimensionality of the problem so that the GPU memory is saturated. Note that for both intra-node and inter-node distributed sampling schemes, the execution times remain nearly constant as long as the number of samples per GPU is kept fixed. This is indicative of near-optimal weak scaling: consider a problem so large that we are able to generate only a few samples using a single GPU due to memory constraints. In this scenario, by using a large number of GPUs to generate independent sets of samples, we should be able to drive the stochastic optimization problem to convergence.

The effective batch size increases as we scale up the number of GPUs. This improves the convergence performance of our method. We benchmark the result in Figure~\ref{fig:scalability_performance}, where we train our models across different numbers of GPUs, on TIM problems of different sizes. The improvement saturates for smaller problems as the effective batch size increases but remains significant for larger problems. This implies that our model requires a larger batch size to achieve optimal performance for problems of a larger scale. Intuitively, batch size quantifies the exploration capability in the state space: the algorithm has a better chance to discover the ground state if it is allowed to explore more.

The raw data of our experiments in this section is provided in the appendix.

\begin{table*}[t]
\small
\caption{Time elapsed to reach the target performance, measured in seconds. We train the RBM with ADAM on Max-Cut problems. At every iteration, after the training updates, we sample another batch of samples for evaluation; the algorithm terminates if the evaluation score surpasses the target score. Evaluation time is not taken into account.}
\label{tbl:benchmark_convergence_time}
\begin{tabular}{ccccccccc}
\toprule
\multirow{2}{*}{Method} &\multicolumn{5}{c}{\# of Dimensions (Targeted cut number)} \\
\cmidrule(lr){2-6}
& 20(41) & 50(190) & 100(730) & 200(2800) & 500(16800) \\
\midrule
MADE+AUTO & 3.14 & 3.61 & 20.08 & 3.25 & 6.27 \\
RBM+MCMC & 126.84 & 154.09 & 247.91 & 612.76 & 1096.08 \\
\toprule
\end{tabular}
\end{table*}

\section{Case Studies}

In this section, we conduct experiments on several aspects of our settings in more detail, to support our conclusions that MADE+AUTO significantly outperforms RBM+MCMC in terms of the convergence rates for large-scale problems. Throughout this section, we train our models for Max-Cut problems with ADAM optimizer on a single GPU.
All results are averaged over 5 runs with different random seeds.

\subsection{Ablation Study: Latent Size}

We conduct ablation studies on the choice of latent size for our models. Latent size refers to the number of hidden units for RBM and the hidden layer size for MADE.

In Table~\ref{tbl:ablation_latent_dim}, we train both MADE and RBM on Max-Cut problems with graph sizes $n \in \{50,100,200,500\}$ under different choices of latent size $h \in \{(\log n)^2,3(\log n)^2,n,5n,n^2\}$. We also cite the numbers from Table~\ref{tbl:benchmark_convergence}
for direct comparison, where we adopt $h=5(\log n)^2,n$ for MADE and RBM, respectively. We measure the training time of each model for 300 iterations in seconds and present the numbers on the right side of the table. The results are averaged over 5 runs with different random seeds.

Several observations can be made. First, optimal performance is obtained under a reasonable choice of $h$, between $3(\log n)^2$ and $n$; models with a latent size that is either too large or too small do not perform well. Second, the time complexity usually does not scale with the model size when running on GPU. However, MADE falls off if we push GPU to its computational limits, \eg, AUTO sampling $bs=1024$ samples from MADE with $O(n^3)$ parameters. This is in practice not a serious concern for MADE with latent size $h=O((\log n)^2)$ as it will always face its memory bottlenecks first by storing the batch of high dimensional inputs as the problem size increases. Third, we re-did the experiments on RBM with $n$ hidden units and obtain slightly different results in Table~\ref{tbl:benchmark_convergence}, due to different choices of random seeds and machines that the model is trained on.

\subsection{Ablation Study: MCMC Sampling Scheme}

We conduct ablation studies on the choice of MCMC sampling schemes. In particular, we consider:
\begin{itemize}
    \item Scheme 1: the sampler discard the first $\{n, 10n\}$ samples in the chain and keep the next $\textit{bs}$ samples.
    \item Scheme 2: the sampler takes every $\{2, 5, 10\}$th sample in the chain until $\textit{bs}$ samples are collected in total.
\end{itemize}

In Table~\ref{tbl:ablation_MCMC_sampling}, we train RBM on Max-Cut problems with graph sizes $n \in \{50,100,200,500\}$ under different choices of MCMC sampling schemes $\{n,10n,\text{\texttimes}2,\text{\texttimes}5,\text{\texttimes}10\}$. We also cite the numbers from Table~\ref{tbl:benchmark_convergence}
for direct comparison, where we discard the first $k$=$3n$+$100$ samples in the MCMC chain. We measure the training time of each model for 300 iterations in seconds and present the numbers on the right side of the table. The results are averaged over 5 runs with different random seeds.

Several observations can be made. First, schemes $10n$ or \texttimes$10$ with longer MCMC chains result in better performance, at the cost of longer running time. Second, when running with GPU, the time complexity only scales with the length of the MCMC chain, but not the $O(n^2)$ model size.

\subsection{Comparison of Hitting Time}

In addition to showing the running time with a fixed number of iterations in Table~\ref{tbl:benchmark_running_time}, we demonstrate that MADE+AUTO also significantly out-performs RBM+MCMC in the sense that the former reach a target performance faster.

In Table~\ref{tbl:benchmark_convergence_time}, we train MADE and RBM on Max-Cut problems with graph sizes $n \in \{50,100,200,500\}$ with target performance $\{41,190,730,2800,16800\}$ that are heuristically chosen based on the results in Table~\ref{tbl:benchmark_convergence}. The performance is measured in seconds and the results are averaged over 5 runs with different random seeds. RBM+MCMC requires a significantly longer time to converge to a target performance for large-scale problems.
\section{Conclusions}

In this work, motivated by recent developments in VQMC made possible by autoregressive sampling, we implemented a distributed variant of VQMC and applied it to solving large-scale quantum systems for which standard random-walk Markov chain Monte Carlo sampling fails to converge.
The main advantage of AUTO compared to MCMC lies in its ability to sample exactly from the distribution of interest, unlike MCMC for which the quality of the generated samples is plagued by unknown convergence time, which becomes a severe problem as the dimension of the problem increases.
Empirically, we demonstrated that AUTO significantly outperforms MCMC in terms of the convergence rates for large-scale problems. Training of AUTO is also more stable than that of MCMC, finding converged solutions that are competitive against the state-of-the-art baselines for Max-Cut.
The above findings motivated us to explore large-scale problems up to 10K dimensions. For that purpose, we built large models and chose a batch size to exhaust the memory usage of each GPU to be distributed. The optimality of our results is only limited by the computational resources available at hand: while
the convergence performance
quickly saturates for small-scale problems, it continues to improve for larger-scale problems as we scale up the number of GPUs.

\section*{Acknowledgements}
Authors gratefully acknowledge support from NSF under grant DMS-2038030. 

\bibliographystyle{ACM-Reference-Format}
\bibliography{references}

\appendix

\section{Raw Data from Multi-GPU Scalability Experiments}

\begin{table*}[t]
\caption{Converged energy and running time for TIM problems of different dimensions. Each GPU is distributed with a batch size of 4; the total batch size equals to 4 times the total number of GPUs used. Paralleling experiments are done across different GPU configurations, where $L_1 \times L_2$ refers to a total $L_1$ number of nodes with $L_2$ GPUs in each node, and the a total number of GPUs is $L=L_1 \times L_2$ . The converged energy improves as the batch size (total number of GPUs) increases.}
\label{tbl:benchmark_scalability_performance}
\begin{tabular}{lcccccccccc}
\toprule
\multirow{2}{*}{\# GPUs} & \multirow{2}{*}{Metric} & \multicolumn{9}{c}{\# of Dimensions} \\
\cmidrule(lr){3-11}
& & 20 & 50 & 100 & 200 & 500 & 1000 & 2000 & 5000 & 10000 \\
\midrule
\midrule
\multirow{2}{*}{$1 \times 1$} & Energy & -69.64 & -225.53 & -656.91 & -1511.22 & -3862.86 & -9642.54 & -21962.55 & -56337.84 & -89733.83 \\
& Time (s) & 2.85 & 5.74 & 10.63 & 20.45 & 49.62 & 98.01 & 204.18 & 514.14 & 1067.56 \\
\midrule
\multirow{2}{*}{$1 \times 2$} & Energy & -70.59 & -260.91 & -626.55 & -1788.10 & -4666.89 & -12056.95 & -24274.07 & -73938.23 & -142214.93 \\
& Time (s) & 3.06 & 6.00 & 10.81 & 20.36 & 49.47 & 97.29 & 200.32 & 512.39 & 1065.71 \\
\midrule
\multirow{2}{*}{$1 \times 4$} & Energy & -82.79 & -257.26 & -702.94 & -1778.35 & -5587.58 & -13797.55 & -29219.47 & -79650.12 & -165364.75 \\
& Time (s) & 3.14 & 6.13 & 10.90 & 20.95 & 49.33 & 98.22 & 202.02 & 507.40 & 1066.03 \\
\midrule
\multirow{2}{*}{$2 \times 2$} & Energy & -82.79 & -257.26 & -702.94 & -1778.35 & -5418.66 & -13286.22 & -28886.57 & -74508.23 & -159416.64 \\
& Time (s) & 3.29 & 6.16 & 10.81 & 20.63 & 49.59 & 98.01 & 204.90 & 512.80 & 1068.00 \\
\midrule
\multirow{2}{*}{$2 \times 4$} & Energy & -81.49 & -261.31 & -766.29 & -1984.61 & -5886.93 & -14826.83 & -31665.81 & -94311.98 & -190800.37 \\
& Time (s) & 5.26 & 7.91 & 11.10 & 20.68 & 49.95 & 100.95 & 206.12 & 515.03 & 1085.33 \\
\midrule
\multirow{2}{*}{$4 \times 2$} & Energy & -81.49 & -261.31 & -766.29 & -1929.95 & -5834.87 & -14464.15 & -33929.40 & -93814.81 & -200729.03 \\
& Time (s) & 3.55 & 6.22 & 10.92 & 20.60 & 49.86 & 97.98 & 202.73 & 513.87 & 1075.07 \\
\midrule
\multirow{2}{*}{$4 \times 4$} & Energy & -81.70 & -261.91 & -776.00 & -1892.16 & -6348.56 & -15636.99 & -44506.68 & -111165.27 & -229567.37 \\
& Time (s) & 3.25 & 6.14 & 13.44 & 21.15 & 49.43 & 98.11 & 203.58 & 514.16 & 1068.51 \\
\midrule
\multirow{2}{*}{$8 \times 2$} & Energy & -81.70 & -261.89 & -776.00 & -1892.15 & -5975.69 & -15928.98 & -46415.26 & -120381.78 & -224738.12 \\
& Time (s) & 3.30 & 6.18 & 10.88 & 20.77 & 49.97 & 98.29 & 203.80 & 520.13 & 1072.32 \\
\midrule
\multirow{2}{*}{$6 \times 4$} & Energy & -80.99 & -276.52 & -769.72 & -1950.40 & -6672.37 & -17105.77 & -38496.40 & -127652.29 & -261517.21 \\
& Time (s) & 3.22 & 6.22 & 11.14 & 21.12 & 50.43 & 101.30 & 206.36 & 521.97 & 1067.83 \\
\toprule
\end{tabular}
\end{table*}

\begin{table*}[t]
\caption{Running time (seconds) for TIM problems of different dimensions. Each GPU is distributed with the maximum number of batchsize that can be accommodated on its memory. A number of differnt GPU configurations were used; $L_1 \times L_2$ indicates $L_1$ nodes with $L_2$ GPUs per node were utilized. We note that for each dimension, the run times remain constant even as we increase the number of GPUs, increasing the effective batch size. This is indicative of near-optimal weak scaling.}
\label{tbl:benchmark_scalability_performance2}
\begin{tabular}{lccccccccc}
\toprule
\multirow{4}{*}{\# GPUs} & \multicolumn{9}{c}{\# of Dimensions} \\
\cmidrule(lr){2-10}
& 20 & 50 & 100 & 200 & 500 & 1000 & 2000 & 5000 & 10000 \\
\cmidrule{2-10}
& \multicolumn{9}{c}{\# of Samples per GPU} \\
\cmidrule(lr){2-10}
& $2^{19}$ & $2^{17}$ & $2^{15}$ & $2^{13}$ & $2^{11}$ & $2^9$ & $2^7$ & $2^4$ & $2^2$ \\
\midrule
$1 \times 1$ & 77.34 & 73.34 & 62.70 & 62.67 & 110.37 & 159.51 & 263.05 & 558.93 & 1058.85 \\
\midrule
$1 \times 2$ & 76.30 & 73.74 & 62.88 & 62.24 & 110.93 & 160.24 & 263.14 & 562.30 & 1060.62 \\
\midrule
$1 \times 4$ & 76.57 & 73.86 & 63.11 & 62.47 & 110.82 & 160.64 & 260.21 & 556.15 & 1054.41 \\
\midrule
$2 \times 2$ & 76.24 & 73.82 & 63.02 & 62.56 & 111.20 & 160.94 & 265.71 & 575.51 & 1068.28 \\
\midrule
$2 \times 4$ & 77.56 & 75.29 & 64.50 & 64.65 & 113.94 & 161.15 & 265.01 & 575.77 & 1075.45 \\
\midrule
$4 \times 2$ & 76.32 & 73.86 & 63.03 & 62.35 & 111.31 & 164.54 & 266.81 & 566.73 & 1070.02 \\
\midrule
$4 \times 4$ & 76.61 & 76.15 & 65.15 & 64.91 & 112.19 & 160.87 & 265.47 & 562.93 & 1071.24 \\
\midrule
$8 \times 2$ & 77.01 & 75.13 & 64.59 & 65.27 & 112.46 & 163.78 & 269.40 & 572.13 & 1077.35 \\
\midrule
$6 \times 4$ & 79.83 & 75.39 & 65.08 & 65.61 & 111.97 & 165.30 & 268.52 & 576.37 & 1073.62 \\
\toprule
\end{tabular}
\end{table*}

We distribute the sampling task across multiple GPUs, our method can extend to large-scale problems with input dimensions up to 10K dimensions, by reducing the mini-batch size $\textit{mbs}$ distributed to each GPU. The effective batch size depends on both $\textit{mbs}$ and the number of GPUs $L$ available for training. Here, we provide the raw data for our distributed computing experiments in Section~\ref{sec:scalability}.

In Table~\ref{tbl:benchmark_scalability_performance}, we show the converged energy and running time for TIM problems of different dimensions. Each GPU is distributed with a batch size of 4; the total batch size equals to 4 times the total number of GPUs used. A number of different GPU configurations were used; $L_1 \times L_2$ indicates $L_1$ nodes with $L_2$ GPUs per node were utilized. The converged energy improves as the batch size (total number of GPUs) increases.

In Table~\ref{tbl:benchmark_scalability_performance2}, we show the running time (seconds) for TIM problems of different dimensions. Different from the experiments in Table~\ref{tbl:benchmark_scalability_performance}, each GPU is distributed with the maximum number of batch size that can be accommodated on its memory.
We note that for each dimension, the run times remain constant even as we increase the number of GPUs, increasing the effective batch size. This is indicative of near-optimal weak scaling.

\end{document}